\documentclass{JAC2003}

\usepackage{graphicx}
\def\beq{\begin{equation}}
\def\eeq{\end{equation}}

\begin{document}

\title{A FORMULA OF THE ELECTRON CLOUD LINEAR MAP COEFFICIENT IN A STRONG DIPOLE}
\author{S. Petracca, A. Stabile, University of Sannio, Benevento (Italy),\\
T. Demma, INFN-LNF, Frascati (Italy),
G. Rumolo, CERN, Geneva (CH)}

\maketitle
\begin{abstract}

Electron cloud effects have recognized as as one of the most serious bottleneck for reaching design performances in presently running and proposed future storage rings. The analysis of these effects is usually performed with very time consuming simulation codes. An alternative analytic approach, based on a cubic map model for the bunch-to-bunch evolution of the electron cloud density, could be useful to determine regions in parameters space compatible with safe machine operations. In this communication we derive a simple approximate formula relating the linear coefficient in the electron cloud density map to the parameters relevant for the electron cloud evolution with particular reference to the LHC dipoles.
\end{abstract}

\section{INTRODUCTION}
In \cite{Iriso} it has been shown that, the evolution of the electron cloud
density can be followed from bunch to bunch introducing a
cubic map of the form:

\beq
\rho_{m+1}\,=\,a\,\rho_m+b\,{\rho_m}^2+c\,{\rho_m}^3
\eeq
where $\rho_l$ is the average line electron density between two
successive bunches, and the coefficients $a$, $b$ and $c$ are extrapolated
from simulations and are function of the beam
parameters and of the beam pipe characteristics.
An analytic expression for the linear map coefficient that describes the particle behavior has been derived from first principles in \cite{Iriso2, epac08}.

\begin{figure}[htb]
\centering
\includegraphics*[width=80mm]{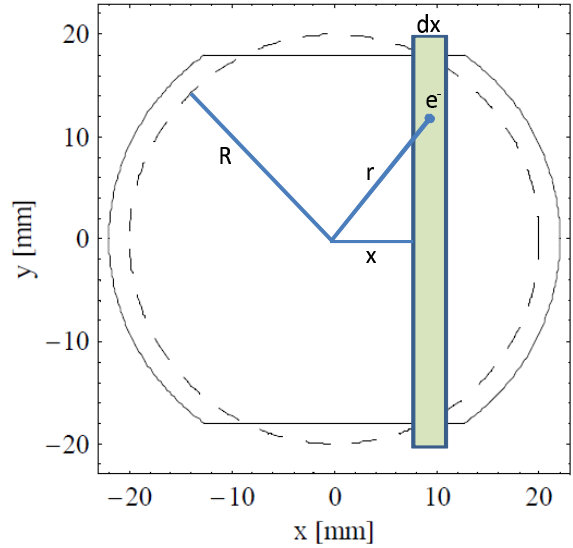}
\caption{Cross section of the LHC screen (solid line) and the circular cross section of the
pipe used in this communication (dashed line).}
	\label{fig:beamscreen}
\end{figure}

In this paper we generalize the model presented in \cite{epac08} in order to take into account the vertical symmetry in the electron cloud distribution induced by the strong vertical magnetic field.
We consider $N_m(x)$ quasi-stationary electrons, where $x$ is the distance from the bunch, uniformly distributed
in a vertical stripe of the transverse cross-section of the beam pipe (Fig. $\ref{fig:beamscreen}$).
The bunch $m$ accelerates the $N_m(x)$ electrons initially at rest to an energy $E_g(x)$.
After the first wall collision two new jets are created: the
backscattered one with energy $E_g$,
and the true secondaries with energy $E_{sec}\,\approx\,5eV$.
The sum over these jets gives the
number of surviving electrons $N_{m+1}$, and the linear coefficient $a$ is obtained by $a\,=\,N_{m+1}/N_m$.

In the next section we compute the electron energy gain $E_g(x)$ due to the passage
of a bunch in the presence of a magnetic dipolar field and compute the number of secondary electrons produced after an electron-wall collision as a function of the electron
energy. Then, following \cite{noi}, we calculate the linear map coefficient $a(x)$ for the case of an LHC-like dipole.

\section{ELECTRON MOTION}

In Fig. $\ref{fig:beamscreen}$ the actual cross section of the
LHC beam-screen together with the circular model used in this paper are shown. In
the presence of a high vertical magnetic field we can consider only the
transverse vertical motion of the electrons.  We approximate the
elicoidal trajectories as vertical straight lines since the cyclotron
radius of the single particle motion is very small with respect to the
transverse beam-pipe radius $R$. In this approximation, the time of flight for an electron with energy $E$ is:

\beq
T_f(E,x)=\sqrt{\frac{2 m_e (R^2 - x^2)}{E}}
\eeq
where $m_e$ is the electron mass.

\begin{figure}[htb]
\centering
\includegraphics*[width=80mm]{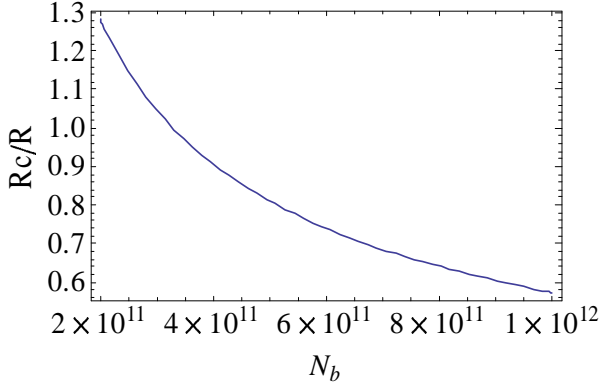}
\caption{Critical radius as a function of bunch population.}
	\label{fig:rc}
\end{figure}

Usually it is possible to distinguish two regimes of the electron cloud. One corresponds to electrons outside the beam core (kick regime), the other to electrons that are trapped within the beam core (autonomous regime) \cite{Berg}.

\begin{figure}[htb]
\centering
\includegraphics*[width=80mm]{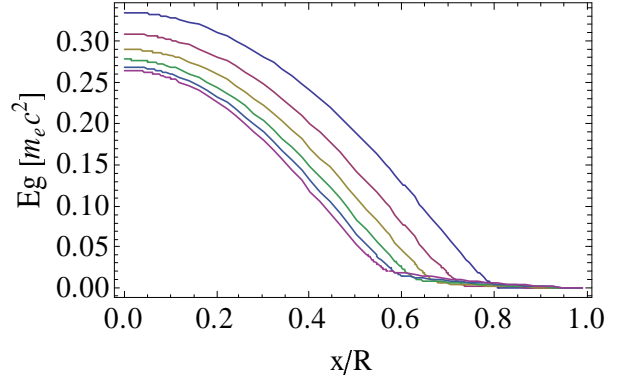}
\caption{Energy gain as a function
of the transverse distance for different values of the bunch population $N_b$.}
	\label{fig_eg}
\end{figure}

The former first one represents the electron motion outside the bunch during a bunch passage, and in the second one the electrons in the beam pipe perform harmonic oscillations. The critical radius $R_c$ separating these two regimes is defined as the radial distance for which the time for the bunch to pass is equal to a quarter of the oscillation period ($R_c\,=\,c\,T/4$). Hence $R_c$ is given as follows

\begin{eqnarray}
R_c\,=\,\frac{\pi}{2}\sqrt{\frac{{\sigma_r}^2\,\sigma_z}{2\,r_e\,N_b}}
\end{eqnarray}
and it is shown in Fig. $\ref{fig:rc}$. The energy gain $E_g(x)$ is evaluated by averaging on the surface of the stripe

\begin{eqnarray}
<E_g>\,=\,S^{-1}\int_S\,dS\,E(x,y)
\end{eqnarray}
where $S$ is the surface of the vertical stripe. Since $R_c$ is of the same order as the considered beam pipe radius, the average energy gain, for $N_m$ electrons uniformly distributed in a vertical stripe of the beam pipe, can be written as:

\begin{eqnarray}
E_g(x)\,=\,\frac{\Theta(x-R_c)}{L}\int_0^{L}dy E_{kick}(x,y)+\,\,\,\,\,\,\,\,\,\,\,\,\,\,\,\,\,\,\,\,\,\,\,\,\,\,\,\,\,\,\,\,
\nonumber\\\frac{\Theta(R_c-x)}{L}\biggl\{\int_{L_c}^{L}dy E_{kick}(x,y)+\int_{-L_c}^{L}dy \frac{E_{aut}(x,y)}{2}\biggr\}
\end{eqnarray}
where

\beq
E_{kick}(x,y)\,=\,m_e c^2 \frac{2{N_b}^{2} {r_e}^2 y^2}{(x^2+y^2)^2}
\eeq

\beq
E_{aut}(x,y)\,=\,m_e c^2\frac{\pi^2\,y^2}{8\,{R_c}^2}
\eeq
$L\,=\,\sqrt{R^2-x^2}$, $L_c\,=\,\sqrt{{R_c}^2-x^2}$, $N_b$ is the bunch population, $r_e$ is the classical electron radius, $\sigma_r$ and $\sigma_z$ are transverse and longitudinal beam dimensions respectively. $\Theta(x)$ is the Heaviside function. The function $E_g(x)$ is shown in Fig. $\ref{fig_eg}$ for different values of the bunch population.

\section{LINEAR MAP}

The electrons multiplication in the beam pipe wall is parameterized by the so-called Secondary Emission Yield (SEY or $\delta$). The total yield is the sum of the"'true secondaries"' $\delta_t(E_g)$ and
"reflected" $\delta_r(E_g)$ electrons.

The $N_m(x)$ electrons gain an energy $E_g(x)$, during the passage of bunch-m, hit the chamber wall and produce $N_m(x) \delta_t$ true secondaries and $N_m(x) \delta_r$ reflected electrons. The reflected electrons travel vertically across the stripe with energy $E_g(x)$ and perform a number of collisions, $n(x)$, with the chamber wall, between two consecutive bunches, that is

\beq
n(x)\,=\,\frac{t_{sb}-T_f(E_g,x)}{T_f(E_g,x)}
\eeq
where $t_{sb}$ is the bunch spacing. Hence, the total number of reflected, high energy electrons at the passage of bunch-$(m+1)$ is

\beq
N_r(x)=N_m(x) {\delta_r(E_g)}^{n(x)}
\eeq
The true secondaries electrons produced after the first wall collision give rise to a low energy jet
($E_{sec} \approx 5 eV$). For this jet there is no distinction between the true secondaries and reflected,
since all are produced with the same energy. After the $i^{th}$ wall collision the number of electrons is

\beq
N_t(x)=N_m(x) \delta_t(E_g) {\delta_r(E_g)}^{i-1} \delta_{sec}^{k_i(x)}
\eeq
where $\delta_{sec}=\delta_r(E_{sec})+\delta_t(E_{sec})$ and

\beq
k_i(x)=\frac{t_{sb}- i T_f(E_{sec},x)}{T_f(E_{sec},x)}
\eeq
is the number of collisions after the  $i^{th}$ collision. The low energy electrons
at the passage of bunch-$(m+1)$ is

\beq
N_s(x)=N_m(x) \delta_t(E_g) \sum_{i=1}^{n(x)} {\delta_r(E_g)}^{i+1} \delta_{sec}^{k_i(x)}
\eeq

\begin{figure}[htb]
\centering
\includegraphics*[width=80mm]{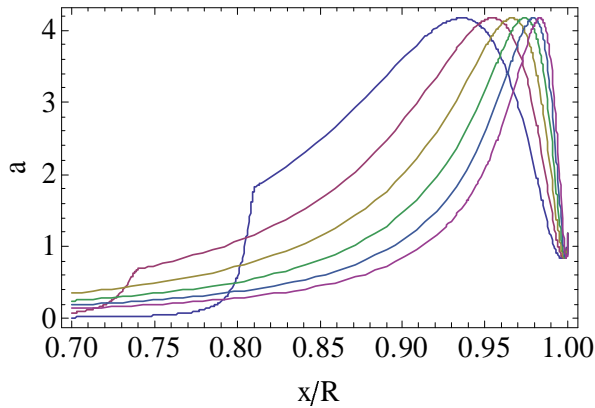}
\caption{Linear map coefficient $a(x)$ for different values of $N_b$.}
	\label{fig:avnb}
\end{figure}

\begin{figure}[htb]
\centering
\includegraphics*[width=80mm]{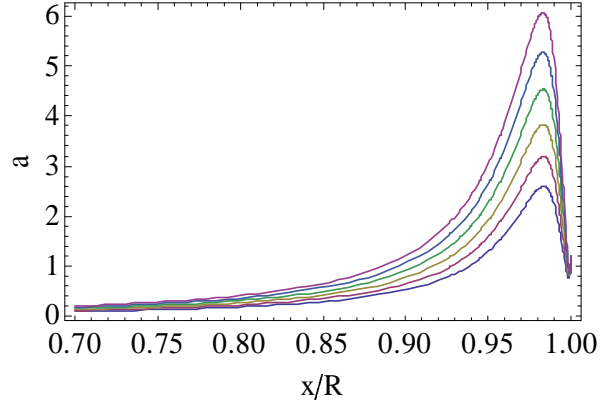}
\caption{Linear map coefficient $a(x)$ for different values of $\delta_{max}$.}
	\label{fig:avSEY}
\end{figure}

Finally the total number of surviving electrons at bunch passage $(m+1)$ is obtained taking into account both the high and low energy contributions

\beq
N_{m+1}(x)=N_m(x) \left( \delta_r^{n(x)} + \delta_t \sum_{i=1}^{n(x)} \delta_r^{i+1} \delta_{sec}^{k_i(x)} \right)
\eeq
and the linear term can be written in the form

\beq
a(x)\,=\,\frac{N_{m+1}(x)}{N_m(x)}
=\delta_r^{n(x)}+\delta_t\delta_{sec}^{\eta}\frac{\delta_{sec}^{n(x) \eta}-\delta_r^{n(x)}}{\delta_{sec}^{\eta}-\delta_r^{n(x)}}
\eeq
where $\eta=\sqrt{E_{sec}/E_g}$. In Fig. $\ref{fig:avnb}$ and $\ref{fig:avSEY}$, respectively, the linear map coefficient is displayed for different values of $N_b$ and $\delta_{max}$.

\begin{table}[hbt]
	\begin{center}
	\caption{Parameters for LHC}
	\begin{tabular}{lcc}

		\hline

		\hline

		parameter& unit & value \\

		\hline

		beam particle energy & GeV & 7000\\

		bunch spacing & m & 7.48\\

		bunch length & m & 0.075\\

		number of  bunches $N_b$ &--&72\\

		number of particles per bunch $N$ & $10^{11}$&1.1 to 10\\

		bending field $B$ &T & 8.4\\

		length of bending magnet &m & 1\\

		vacuum screen half height & m& 0.018\\

		vacuum screen half width & m& 0.022\\

		circumference &m & 27000\\

		primary photo-emission yield &-&$7.98\cdot10^{-4}$\\

    maximum $SEY$ $\delta_{max}$ &-&1.3 to 2.5\\

		energy for max. $SEY$ $E_{max}$ & eV & 237.125\\

		energy width for secondary $e^-$ & eV &1.8\\

		\hline

		\hline

	\end{tabular}
\label{tab1}
\end{center}
\end{table}

\section{CONCLUSIONS}
The linear map coefficient is analytically derived for the evolution of an electron cloud density inside a dipole.
The analysis is useful to determine safe regions in parameter space in order to reduce the electron cloud effects.


\begin{thebibliography}{9}

\bibitem{Iriso}
U.Iriso and S.Peggs, "Maps for Electron Clouds", Phys.Rev. ST-AB8, 024403, 2005.

\bibitem{Iriso2}
U. Iriso and S. Pegg, Proc. of EPAC06, pp. 357-359.

\bibitem{epac08}
T. Demma and S. Petracca, Proc. of EPAC08, pp. 1601-1603.

\bibitem{noi}	
T.Demma et al., "Maps for Electron Clouds: Application To LHC", Phys.Rev.ST-AB10, 114401 (2007).

\bibitem{Berg}
J. Scott Berg, "Energy gain in an electron cloud during the passage of a bunch", LHC Project Note 97, CERN, Geneva (CH), 1997.

\bibitem{ecloud}
http://wwwslap.cern.ch/electron-cloud/Programs/Ecloud/\\ecloud.html.

\end{thebibliography}
\end{document}